\documentstyle[aps,prl,floats,epsfig]{revtex}

\newcommand{\bastar}{\begin{eqnarray*}}
\newcommand{\eastar}{\end{eqnarray*}}
\newskip\humongous \humongous=0pt plus 1000pt minus 1000pt

\newif\ifdtup

\relax
\newcommand{\be}{\begin{equation}}
\newcommand{\ee}{\end{equation}}
\newcommand{\bea}{\begin{eqnarray}}
\newcommand{\eea}{\end{eqnarray}}

\newcommand{\tF}{{\tilde F}}

\newcommand{\dfrac}{\displaystyle\frac}
\newcommand{\ba}{\begin{array}}
\newcommand{\ea}{\end{array}}

\newcommand{\nn}{\nonumber}

\begin{document}
\twocolumn[\hsize\textwidth\columnwidth\hsize\csname@twocolumnfalse%
\endcsname
\title  {Effective Action---A Convergent Series---of QED}
\bigskip

\author{Y. M. Cho$^{1, 2}$ and D. G. Pak$^{2
}$}

\address{
$^{1)}$Department of Physics, College of Natural Sciences, Seoul National University,
Seoul 151-742, Korea  \\
$^{2)}$Asia Pacific Center for Theoretical Physics, 207-43 Cheongryangri-dong, Dongdaemun-gu,
               Seoul 130-012 Korea\\
{\scriptsize \bf ymcho@yongmin.snu.ac.kr,
dmipak@mail.apctp.org} \\ \vskip 0.3cm
}
\maketitle

\begin{abstract}

The one-loop effective action of QED
obtained by Euler and Heisenberg
and by Schwinger has been  expressed by an asymptotic perturbative series which is divergent.
In this letter we present a non-perturbative but convergent series of
the effective action. With the convergent series we establish
the existence of the manifest electric-magnetic duality in
the one loop effective action of QED.

\vspace{0.3cm}
\end{abstract}

\narrowtext
\bigskip
                           ]
It has been well known that Maxwell's electrodynamics
gets a quantum correction due to
the electron loops. This quantum correction has first
been studied by Euler and Heisenberg
 and  by  Schwinger long time ago
 \cite{euler,schw1}, and later by
many others in detail \cite{ditt,ritus}.
The physics behind the quantum correction is also very
well understood, and the various non-linear effects
arising from the quantum corrections (the pair production,
the vacuum birefringence, the photon splitting, etc.) are being
tested and confirmed by experiments
\cite{exp,bial}.

Unfortunately it is also very well known that the one-loop effective action
of QED has been  expressed only by a perturbative  series which is divergent.
For example, for a uniform magnetic
field $B$, the Euler-Heisenberg effective action is given by \cite{schw1,ditt}
\bea
\Delta {\cal L} &\simeq& -\dfrac{2m^4}{\pi^2}  (\dfrac{eB}{m^2})^4 \nn  \\
 &\times & {\sum_{n=0}^{\infty}}
  \dfrac{2^{2n} B_{2 n +4} }{(2n+2)(2n+3)(2n+4)} (\dfrac{eB}{m^2})^{2n},
\eea
where $m$ is the electron mass and $B_n$ is the Bernoulli number. Clearly
the series (1) is an asymptotic series which is divergent  \cite{ditt,ritus}.
This is not surprising. In fact one could argue that
the effective action, as a perturbative series,
can only be expressed by a divergent asymptotic series \cite{dyson,zhit}.
This suggests that only a non-perturbative series could provide
a convergent expression for the effective action. There have been many
attempts to improve the convergence of the series with a Borel-Pade
resummation \cite{zhit,dunne}. Although these attempts have made remarkable
progresses for various purposes, they have not
produced a convergent series so far.

The purpose
of this letter is to provide a non-perturbative but convergent series of
the one-loop effective action of QED.
{\it
Using a non-perturbative series expansion we prove that
the one loop effective action of QED
can be expressed by }
\bea
&&{\cal L}_{eff} = -\dfrac{a^2 - b^2}{2}(1-\dfrac{e^2}{12\pi^2} 
\ln\dfrac{m^2}{\mu^2}) \nn\\
&&- \dfrac{e^2 }{4\pi^3} ab  \sum_{n=1}^{\infty}\dfrac{1}{n}
{\Big [}\coth(\dfrac{n \pi b}{a}){\Big (} {\rm ci}(\dfrac{n \pi m^2}{ea}) 
\cos(\dfrac{n \pi m^2}{ea}) \nn\\
&&+{\rm si}(\dfrac{n \pi m^2}{ea}) \sin(\dfrac{n \pi m^2}{ea}){\Big )}\nn\\
&&- \dfrac{1}{2} \coth ( \dfrac{n \pi a}{b})  {\Big (} \exp(\dfrac{n \pi m^2}{eb})
{\rm Ei}(-\dfrac{n \pi m^2}{eb} )   \nn \\
&&+ \exp(-\dfrac{n \pi m^2}{eb} ) {\rm Ei}(\dfrac{n \pi m^2}{eb} ){\Big )}{\Big ]},
\eea
where $\mu$ is the subtraction parameter and
\bea
a = \dfrac{1}{2} \sqrt {\sqrt {F^4 + (F \tF)^2} + F^2}, \nn \\
b = \dfrac{1}{2} \sqrt {\sqrt {F^4 + (F \tF)^2} - F^2}. \nn
\eea
Clearly the series is not perturbative,
but convergent. The series expression
has a great advantage over the divergent perturbative
series. It allows us to have
a massless limit. Furthermore it
has a manifest electric-magnetic duality, as we will
discuss in the following.

For the scalar QED we also obtain a similar convergent series
for the effective action which has a smooth massless limit and the
manifest duality. Our results become important when we evaluate the
effective action of QCD.

To derive the effective action let's start from the QED Lagrangian
\bea
{\cal L} &=& -\dfrac{1}{4} F_{\mu \nu}^2 + \bar \Psi( i {{/ \,} \llap D} -m) \Psi .
\eea
With a proper gauge fixing
one can show that one loop fermion correction of the effective action is given by
\bea
\Delta S &=& i \ln {\rm Det} (i {{/ \,} \llap D}  - m) .
\eea
So for an arbitrary constant background
one has
\bea
\Delta {\cal L} = - \dfrac{e^2}{8\pi^2} ab \int_{0}^{\infty} \dfrac{dt}{t}
\coth (eat) \cot (ebt) e^{-m^2 t}.
\eea
This integral expression, of course, has been known for a long
time \cite{schw1,ditt}. However, as far as we understand, the integral
has been performed only in a perturbative series which is divergent
(except for the special cases of $a$ and $b$) \cite{dunne,miel}.

To perform the integral one must choose the contour of the integral
first. Here the causality dictates the contour to pass above the real axis.
Now, to obtain a convergent series of the integral the following
Sitaramachandrarao's identity
plays the crucial role \cite{sita},
\bea
& xy \coth x \cot y = 1 + \dfrac{1}{3} (x^2-y^2)  \nn \\
& -  \dfrac{2}{\pi}  x^3 y  \sum_{n=1}^{\infty} \dfrac{1}{n} \dfrac{\coth
(\dfrac{n \pi y}{x})}{(x^2 +n^2  \pi^2)}  \nn \\
&  +\dfrac{2}{\pi}    x y^3  \sum_{n=1}^{\infty} \dfrac{1}{n}
 \dfrac{\coth (\dfrac{n \pi x}{y})}{(y^2 - n^2 \pi^2)}.
\eea
With the identity we have
\bea
\Delta {\cal L} = I_1(\epsilon , m) + I_2 (\epsilon , m)+ I_3(\epsilon , m),
\eea
where $\epsilon$
is the ultra-violet cutoff parameter and
\bea
I_1 &=& -\dfrac{1}{8 \pi^2} \int_0^{\infty} t^{\epsilon -3} ( 1 + e^2\dfrac{a^2 - b^2}{3}t^2 ) e ^{-m^2 t} dt \nn \\
\,\, & \simeq& -\dfrac{1}{8 \pi^2} {\Big [}( \dfrac{m^4}{2} + e^2\dfrac{a^2 - b^2 }{3 })
             (\dfrac{1}{\epsilon} - \gamma -\ln \dfrac{m^2}{\mu^2} ) + \dfrac{3}{4} m^4 {\Big ]}, \nn \\
I_2 &=& \dfrac{e^2}{4 \pi^3} ab \sum_{n=1}^{\infty}
\dfrac{
     \coth (
             \dfrac{n \pi b}{a}
                                 )
                                    } {n}
 \int_0^{\infty} \dfrac{t^{\epsilon +1} e^{-m^2 t} }
{t^2 + (\dfrac{n \pi}{ea})^2} dt \nn \\
&\simeq& -\dfrac{e^2}{4 \pi^3} ab \sum_{n=1}^{\infty} \dfrac{\coth ( \dfrac{n \pi b}{a})}{n}
{\Big [}{\rm ci}
(\dfrac{n \pi m^2 }{ea}) \cos  (\dfrac{n \pi m^2 }{ea})   \nn  \\
&\, & + {\rm si} (\dfrac{n \pi m^2 }{ea}) \sin(\dfrac{n \pi m^2 }{ea}){\Big ]}, \nn \\
 I_3 &=&-  \dfrac{e^2 }{4 \pi^3} ab  \sum_{n=1}^{\infty} \dfrac{\coth ( \dfrac{n \pi a}{b})}{n}
 \int_0^{\infty} \dfrac{t^{\epsilon +1} e^{-m^2 t} }
{t^2 - (\dfrac{n \pi}{eb})^2} dt \nn \\
&\simeq& \dfrac{e^2}{8 \pi^3} ab \sum_{n=1}^{\infty} \dfrac{\coth ( \dfrac{n \pi a}{b})}{n}
{\Big [} \exp (\dfrac{n \pi m^2}{eb}) {\rm Ei}  (-\dfrac{n \pi m^2}{eb})   \nn \\
&\, &
+\exp (-\dfrac{n \pi m^2}{eb}) {\rm Ei } (\dfrac{n \pi m^2}{eb}){\Big ]} .
\eea
So with
the ultra-violet regularization by the modified minimal subtraction
we obtain the convergent series expression (2), where we have neglected 
the (irrelevant) cosmological constant term. Observe the appearance
of the logarithmic correction in the classical part of the action, 
which plays an important role in
the discussion of the renormalization invariance of the 
effective action. 

The effective action has an imaginary part
when $b \neq 0$,
\bea
{\rm Im}\, \Delta {\cal L} =  \dfrac{e^2}{8 \pi^2 } ab  \sum_{n=1}^{\infty}
    \dfrac{1}{n}   \coth (\dfrac{n \pi a}{b})\, \exp(-\dfrac{n \pi m^2}{eb}).
\eea
This is because the exponential integral
${\rm Ei}(-z)$ in (8) develops an imaginary part after the analytic continuation
from $z$ to $-z$. The important point here is that the analytic
continuation should  be made in such a way to preserve the causality,
which determines the signature of the imaginary part in (9).
The physical meaning of the imaginary part is well known \cite{schw1}.
The electric background generates the pair creation,
with the probability per unit volume per unit time given by (9).

Clearly
our series expression has a great advantage
over the asymptotic series. An immediate advantage
is that it naturally  allows a massless limit. To see this notice that in
the massless limit we have
\bea
 I_1 &\simeq& -\dfrac{e^2  }{24 \pi^2} (a^2 - b^2) (\dfrac{1}{\epsilon} -
             \gamma - \ln \dfrac{m^2}{\mu^2}), \nn \\
  I_2 &\simeq& -\dfrac{e^2}{4 \pi^3} ab  \sum_{n=1}^{\infty} \dfrac{\coth (\dfrac{n \pi b}{a})}{n}
       {\Big (}\gamma
     + \ln (\dfrac {n \pi \mu^2}{ea}) + \ln \dfrac {m^2}{\mu^2} {\Big )}, \nn \\
  I_3 &\simeq & \dfrac {e^2 }{4 \pi^3} ab  \sum_{n=1}^{\infty} \dfrac{\coth (\dfrac{n \pi a}{b})}{n}
       {\Big (}\gamma
         + \ln (\dfrac {n \pi \mu^2}{eb}) + \ln \dfrac {m^2}{\mu^2} \nn \\
&& + i \dfrac{\pi}{2}  {\Big )},
\eea
so that
\bea
&& \Delta {\cal L} \simeq  -\dfrac{e^2 }{24 \pi^2}(a^2 - b^2) (\dfrac{1}{\epsilon} - \gamma ) +     \dfrac{e^2}{24 \pi^2} {\Big [} a^2 - b^2 \nn \\
&& - \dfrac{6 ab }{\pi} \sum_{n=1}^{\infty}
  \dfrac{1}{n} {\Big (}\coth (\dfrac{n \pi b}{a}) -\coth  (\dfrac{n \pi a}{b}){\Big )}
                {\Big ]} \ln \dfrac{m^2 }{\mu^2}  \nn \\
&&  - \dfrac{e^2 }{ 4 \pi^3}ab \sum_{n=1}^{\infty}
  \dfrac{1}{n} {\Big [} \coth (\dfrac {n \pi b}{a}) {\Big (}
\gamma + \ln ( \dfrac{n \pi \mu^2}{ea}){\Big )} - \coth  (\dfrac{n \pi a}{b}) \nn \\
&&  {\Big (} \gamma + \ln ( \dfrac{n \pi \mu^2}{eb}){\Big )}
{\Big ]}
            + i \dfrac{e^2}{8 \pi^2 }ab \sum_{n=1}^{\infty} \dfrac{1}{n} \coth (\dfrac{n \pi a}{b}) .
\eea
From this we can separate the finite part of the effective
action from the infra-red
divergence. For $ab \neq 0$ we obtain (with the modified minimal
ultra-violet subtraction)
\bea
 \Delta {\cal L}{\Big |}_{m=0} = \Delta {\cal L}_{\infty} + \Delta {\cal L}_{fin},
\eea
where
\bea
 &&  \Delta {\cal L}_{\infty}
       \simeq \dfrac{e^2 }{24 \pi^2} {\Big [} a^2 - b^2 -
     \dfrac{6 a b }{\pi} \sum_{n=1}^{\infty}  \dfrac{1}{n} {\Big (} \coth (\dfrac {n \pi b }{a}) \nn \\
 &&-  \coth (\dfrac {n \pi a }{b}) {\Big )} {\Big ]}  \ln \dfrac{m^2}{\mu^2}
        + \dfrac{e^2}{8 \pi^3} a b \sum_{n=1}^{\infty}  \dfrac{1}{n} {\Big (} \coth (\dfrac {n \pi b }{a}) \nn \\
 &&+\coth (\dfrac {n \pi a }{b}) {\Big )} \ln \dfrac{a}{b}
   + i \dfrac{e^2}{8 \pi^2} a b \sum_{n=1}^{\infty}  \dfrac{1}{n}
           \coth (\dfrac{n \pi a}{b}),
\eea
and
\bea
   \Delta {\cal L}_{fin} =&&-\dfrac{e^2}{8 \pi^3} a b
                    \sum_{n=1}^{\infty}  \dfrac{1}{n} {\Big (} \coth (\dfrac {n \pi b }{a})
             -  \coth (\dfrac {n \pi a }{b}) {\Big )} \nn \\
&&   {\Big (} 2 \gamma + \ln (\dfrac{n \pi \mu^2}{ea}) +
 \ln (\dfrac {n \pi \mu^2}{eb}) {\Big )}.
\eea
Clearly this separation of the infra-red divergence
was not possible with the old asymptotic series.

An important point here is that the logarithmic infra-red divergence
in (11) disappears when (and only when) $ab=0$, due to the identity
\bea
 \dfrac{6}{\pi} ab \sum_{n=1}^{\infty}
  \dfrac{1}{n} {\Big (}\coth (\dfrac{n \pi b}{a})
-\coth (\dfrac{ n \pi a}{b}){\Big )} = a^2 - b^2.
\eea
Furthermore in this case the remaining part of (11) becomes
finite (after the ultra-violet subtraction).
Indeed one finds
\bea
\Delta {\cal L} {\Big |}_{m=0} = \left\{{~~\dfrac{e^2a^2 }
{24 \pi^2} (\ln \dfrac{ea}{\mu^2} -c)~~~~~~~~~~~~~~ b=0, 
\atop -\dfrac{e^2b^2 }{24 \pi^2} (\ln \dfrac{eb}{\mu^2} -c)
+ i \dfrac{e^2b^2}{48 \pi} ~~~a=0,}\right.
\eea
where
\bea
c = \gamma + \ln \pi + \dfrac{6}{\pi^2} \sum_{n=1}^{\infty}  \dfrac{1}{n^2} \ln n = 2.2919...
\eea
This shows that, when $ab=0$, the effective action of QED
does not have any infra-red divergence even in the massless limit.
This agrees with the known result \cite{ditt,ritus}.

{\it A remarkable feature of our effective action is that it is
manifestly invariant under the dual transformation,}
\bea
a \rightarrow -ib,~~~~b \rightarrow  ia.
\eea
{\it
This tells that, as a function of $z=a+ib$, the effective action is
invariant under the reflection from $z$ to $-z$}.  Notice that, in the
Lorentz frame where $\vec E$ is parallel to $\vec B$, $a$ becomes
$B$ and $b$ becomes $E$. So the duality describes the electric-magnetic
duality. To prove the duality notice that the dual transformation
automatically involves the analytic continuation of  
the special functions ci($x$), si($x$), and Ei($x$) in (2). 
With the correct analytic continuation 
we can establish the duality in our effective action.
One might think that the duality is obvious since it immediately follows from 
the integral expression (5). This is not so. In fact the integral
expression is invariant under the four transformations,
\bea
a \rightarrow \pm ~ib,~~~~b \rightarrow  \pm ~/\mp ia.
\eea
But among the four only our 
duality (18) survives as the true
symmetry of the effective action. So the duality constitutes
a non-trivial symmetry of the quantum effective action.
From the physical point of view the existence of the duality 
in the effective action
is perhaps not so surprising. But the fact that this
duality is borne out from our calculation of one loop effective action 
is really remarkable.

One can obtain the similar results for the scalar QED. In this case
the one loop correction is given by \cite{schw1,ditt}
\bea
\Delta {\cal L}_0 = \dfrac {e^2}{16 \pi^2}ab
\int_0^\infty \dfrac{dt}{t}  {\rm csch} (eat) {\rm csc} (ebt)
           e^{- m^2 t} .
\eea
To perform the integral we introduce a new identity similar to the
Sitaramachandrarao's identity (6)
\bea
& xy {\rm csch} x \csc y  = 1 - \dfrac{1}{6} (x^2 - y^2)  \nn \\
& - \dfrac{2}{\pi} x^3 y  \sum_{n=1}^\infty \dfrac{(-1)^n}{n} \dfrac{{\rm csch}
(\dfrac{n \pi y}{x})}
             {x^2 + n^2 \pi^2 } \nn \\
& + \dfrac{2}{\pi} x y^3 \sum_{n=1}^\infty \dfrac{(-1)^n}{n}
\dfrac{{\rm csch} (\dfrac{n \pi x}{y})}
             {y^2 - n^2 \pi^2 },
\eea
and obtain
\bea
\Delta {\cal L}_0 = J_{1}(\epsilon , m) + J_2 (\epsilon , m)+ J_3(\epsilon , m),
\eea
where
\bea
J_1  &\simeq& \dfrac{1}{16 \pi^2} {\Big [}( \dfrac{m^4}{2} - e^2  \dfrac{a^2 - b^2 }{6 })
(\dfrac{1}{\epsilon} - \gamma - \ln \dfrac{m^2}{\mu^2}  )
+ \dfrac{3}{4} m^4 {\Big]}, \nn \\
J_2
&\simeq& \dfrac{e^2}{8 \pi^3} ab \sum_{n=1}^{\infty}
      \dfrac{(-1)^n}{n} {\rm csch} (\dfrac{n \pi b}{a})   \nn \\
&& {\Big [}{\rm ci}
(\dfrac{n \pi m^2 }{ea}) \cos  (\dfrac{n \pi m^2 }{ea})
 + {\rm si} (\dfrac{n \pi m^2 }{ea}) \sin(\dfrac{n \pi m^2 }{ea}){\Big ]}, \nn \\
 J_3
&\simeq& -\dfrac{e^2}{16 \pi^3} ab \sum_{n=1}^{\infty} \dfrac{(-1)^n}{n}
 {\rm csch} (\dfrac{n \pi a}{b})  {\Big [} \exp (\dfrac{n \pi m^2}{eb}) \nn \\
 &&  {\rm Ei}  (-\dfrac{n \pi m^2}{eb})
+\exp (-\dfrac{n \pi m^2}{eb}) {\rm Ei } (\dfrac{n \pi m^2}{eb}){\Big ]} .
\eea
With this we finally obtain with the modified minimal subtraction
(again neglecting the cosmological term)
\bea
&&{\cal L}_{0\,eff}= -\dfrac{a^2 - b^2}{2}(1-\dfrac{e^2}{48\pi^2} 
\ln\dfrac{m^2}{\mu^2}) \nn\\
&&+ \dfrac{e^2}{8\pi^3} ab \sum_{n=1}^{\infty}\dfrac{(-1)^n}{n}
{\Big [}{\rm csch} (\dfrac{n \pi b}{a}) {\Big (} {\rm ci}
(\dfrac{n \pi m^2 }{ea}) \cos (\dfrac{n \pi m^2 }{ea}) \nn\\
&&+ {\rm si} (\dfrac{n \pi m^2 }{ea}) \sin(\dfrac{n \pi m^2 }{ea}) {\Big )}\nn\\
&&-\dfrac{1}{2}  {\rm csch}(\dfrac{n \pi a}{b})
{\Big (} \exp (\dfrac{n \pi m^2}{eb}) {\rm Ei}(-\dfrac{n \pi m^2}{eb})\nn \\
&&+\exp(-\dfrac{n \pi m^2}{eb}){\rm Ei} (\dfrac{n \pi m^2}{eb}) {\Big )} {\Big ]}.
\eea
Again notice the appearance of the logarithmic correction.
The effective action develops an imaginary part,
\bea
 & {\rm Im} \Delta {\cal L}_0 =
          - \dfrac{e^2}{16 \pi^2 }ab \sum_{n=1}^{\infty}
  \dfrac{ (-1)^n}{n}  {\rm csch} (\dfrac{n \pi a}{b}) \nn \\
&  \exp (-\dfrac{n \pi m^2}{eb}).
\eea
Observe that the effective action of the scalar QED
also has the manifest duality.

The  effective action of the scalar QED has a smooth massless limit.
For $m \simeq 0$
we have
\bea
&&    \Delta {\cal L}_0 \simeq \dfrac{e^2}{96 \pi^2}{\Big [}  a^2 - b^2 +
     \dfrac{12 ab  }{ \pi} \sum_{n=1}^{\infty}
     \dfrac{(-1)^n}{n} {\Big (} {\rm csch} (\dfrac{n \pi b}{a})  \nn \\
&&   - {\rm csch}  (\dfrac{n \pi a}{b}){\Big )} {\Big ]} \ln \dfrac{m^2 }{\mu^2}
         + \dfrac{e^2 }{ 8 \pi^3}ab \sum_{n=1}^{\infty}
        \dfrac{(-1)^n}{n} {\Big [}{\rm csch} (\dfrac{n \pi b}{a}) \nn \\
&&        {\Big (}\gamma +
        \ln (\dfrac{ n \pi \mu^2}{ea}){\Big )}  -  {\rm csch}
(\dfrac{n \pi a}{b}) {\Big (} \gamma
        + \ln (\dfrac{ n \pi \mu^2}{eb}){\Big )}{\Big ]}  \nn \\
&&              - i \dfrac{e^2}{16 \pi^2 }ab \sum_{n=1}^{\infty}
         \dfrac{ (-1)^n}{n}  {\rm csch} (\dfrac{n \pi a}{b}) .
\eea
But remarkably the logarithmic infra-red divergence disappears completely due
to the following identity
\bea
& \dfrac{12}{\pi} ab \sum_{n=1}^{\infty} \dfrac{(-1)^n}{n} {\Big (}
 {\rm csch} (\dfrac{n \pi b}{a}) - {\rm csch}
(\dfrac {n \pi a}{b}){\Big )} \nn \\
& = b^2 - a^2.
\eea
Furthermore the remaining part of (25) becomes finite,
\bea
&\Delta {\cal L}_0 {\Big |}_{m =0} = \dfrac{e^2 }{ 8 \pi^3}ab \sum_{n=1}^{\infty}
  \dfrac{(-1)^n}{n} {\Big [} {\rm csch} (\dfrac{n \pi b}{a})  \nn \\
&    {\Big (} \gamma + \ln (\dfrac{ n \pi \mu^2}{ea}){\Big )}
 -{\rm csch} (\dfrac{n \pi a}{b}) {\Big (}\gamma
+ \ln ( \dfrac{n \pi \mu^2}{eb}){\Big )} {\Big ]} \nn \\
&  - i \dfrac{e^2}{16 \pi^2 }ab \sum_{n=1}^{\infty}
  \dfrac{ (-1)^n}{n}  {\rm csch} (\dfrac{n \pi a}{b}).
\eea
{\it This shows that our series expression of the scalar QED
does not contain any infra-red divergence in the massless limit,
even when $ab \neq 0$}. This is really
remarkable, which should be contrasted with the real QED which has a genuine
infra-red divergence when $ab \neq 0$.

Clearly our result should become very useful in studying the non-linear effects
of QED. More importantly our effective action provides a new method
to estimate the running coupling constant non-perturbatively. 
This (and the renormalization of the effective action) will
be discussed in the succeeding paper \cite{cho}.

One of the authors (YMC) thanks Professor S. Adler, Professor F. Dyson,
and Professor C. N. Yang for the illuminating discussions.
The work is supported in part by
Korea Research Foundation (KRF-2000-015-BP0072), and by
the BK21 project of Ministry of Education.

\end{document}